\documentclass[useAMS,usenatbib]{mn2e}

\usepackage{amssymb}
\usepackage{graphicx}
\usepackage{url}

\def\aj{AJ}
\def\apj{ApJ}
\def\apjl{ApJ}
\def\apjs{ApJS}
\def\aap{A\&A}
\def\aaps{A\&AS}
\def\mnras{MNRAS}
\def\pasa{PASA}
\def\nat{Nature}

\newcommand\Msun{\,\rmn{M}_\odot}
\newcommand\ML{\,(\rmn{M/L})_\odot}

\title[Stripped nuclei in galaxy clusters]{Contribution of stripped nuclear clusters to globular cluster and ultra-compact dwarf galaxy populations}
\author[J. Pfeffer et al.]{J. Pfeffer$^{1,2}$\thanks{E-mail: j.pfeffer@uq.edu.au}, B.~F. Griffen$^{3}$, H. Baumgardt$^{1}$ and M. Hilker$^{2}$ \\
$^{1}$School of Mathematics and Physics, The University of Queensland, Brisbane, QLD 4072, Australia \\
$^{2}$European Southern Observatory (ESO), Karl-Schwarzschild-Strasse 2, 85748 Garching, Germany \\
$^{3}$Massachusetts Institute of Technology, Kavli Institute for Astrophysics and Space Research, 77 Massachusetts Avenue, Cambridge, \\ MA 02139, USA}

\begin{document}

\date{}

\pagerange{\pageref{firstpage}--\pageref{lastpage}} \pubyear{2014}

\maketitle

\label{firstpage}

\begin{abstract}
We use the Millennium II cosmological simulation combined with the semi-analytic galaxy formation model of \citet{Guo:2011} to predict the contribution of galactic nuclei formed by the tidal stripping of nucleated dwarf galaxies to globular cluster (GC) and ultra-compact dwarf galaxy (UCD) populations of galaxies. We follow the merger trees of galaxies in clusters back in time and determine the absolute number and stellar masses of disrupted galaxies. We assume that at all times nuclei have a distribution in nucleus-to-galaxy mass and nucleation fraction of galaxies similar to that observed in the present day universe. Our results show stripped nuclei follow a mass function $N(M) \sim M^{-1.5}$ in the mass range $10^6 < M/\Msun < 10^8$, significantly flatter than found for globular clusters. The contribution of stripped nuclei will therefore be most important among high-mass GCs and UCDs. For the Milky Way we predict between 1 and 3 star clusters more massive than $10^5 \Msun$ come from tidally disrupted dwarf galaxies, with the most massive cluster formed having a typical mass of a few times $10^6 \Msun$, like omega Centauri. For a galaxy cluster with a mass $7 \times 10^{13} \Msun$, similar to Fornax, we predict $\sim$19 UCDs more massive than $2\times10^6 \Msun$ and $\sim$9 UCDs more massive than $10^7\Msun$ within a projected distance of 300 kpc come from tidally stripped dwarf galaxies. The observed number of UCDs are $\sim$200 and 23, respectively. We conclude that most UCDs in galaxy clusters are probably simply the high mass end of the GC mass function.
\end{abstract}


\section{INTRODUCTION}

Ultra-compact dwarf galaxies (UCDs) are a class of stellar systems that was discovered more than a decade ago in spectroscopic surveys of the Fornax cluster \citep*{Hilker:1999,Drinkwater:2000}. They have since been discovered in other galaxy clusters \citep*{Mieske:2004, Hasegan:2005, Jones:2006, Mieske:2007, Misgeld:2011, Madrid:2011, Penny:2012}, galaxy groups \citep{Evstigneeva:2007b, DaRocha:2011}, as well as isolated spiral galaxies \citep{Hau:2009}. UCDs have typical ages of 10 to 11 Gyr \citep{Evstigneeva:2007a,Francis:2012} and are typically defined to have half-light radii $7 \lesssim r_h / \rmn{pc} \lesssim 100$ and masses $M \gtrsim 2 \times 10^6 \Msun$, making them an intermediate object between globular clusters (GCs) and dwarf galaxies.

The exact formation mechanism of UCDs and their relation to GCs and dwarf galaxies is unknown and under much debate, although a number of scenarios have been proposed. The simplest explanation is that they are the high-mass end of the GC mass function observed around galaxies with rich GC systems \citep*{Mieske:2002, Mieske:2012} where a physical mechanism causes an increasing lower size limit with increasing mass \cite[e.g.][]{Murray:2009}. Since UCDs have larger sizes than typical GCs, they may be formed from the merger of many GCs in star cluster complexes \citep{Kroupa:1998, Fellhauer:2002, Bruens:2011, Bruens:2012}. Alternatively, they could be nucleated dwarf galaxies stripped by tidal interactions such that only their compact central region remains, referred to as the tidal stripping or `threshing' scenario \citep*{Bassino:1994, Bekki:2001, Bekki:2003, Drinkwater:2003, Pfeffer:2013}. There is also evidence suggesting UCDs are a mix of stellar systems from different formation scenarios rather than a single one \citep{Mieske:2006, Brodie:2011, Chilingarian:2011, DaRocha:2011, Norris:2011}.

In general, it is impossible to determine the formation mechanism of a given UCD because the predictions of internal UCD properties (size, mass, colour, velocity dispersion and metallicity) are very similar between the formation scenarios \citep[although there are exceptions, such as NGC 4546 UCD1 which is evidently the result of tidal stripping,][]{Norris:2011}. Given this, comparison of UCD ages, numbers and spatial distributions with those of genuine GCs and dwarf galaxies are needed to determine the relative importance of the different UCD formation scenarios.

In this work, we concentrate on the tidal stripping scenario. Tidal stripping of nucleated dwarf galaxies is a likely origin for at least some part of the UCD population for a number of reasons: Observational studies show UCDs and early-type galaxy nuclei have many common properties \citep{Evstigneeva:2008, Brodie:2011}. Theoretical work has also demonstrated that the sizes, masses and internal velocity dispersions of stripped nuclei and UCDs are very similar \citep{Bekki:2003, Pfeffer:2013}. Irregular objects with asymmetric extensions have been found which may be dwarf galaxy nuclei undergoing tidal stripping \citep{Richtler:2005, Brodie:2011}. Current theories of giant elliptical formation in galaxy clusters suggest the dominant growth mechanism for the galaxies from $z=1$ to $z=0$ is accretion through minor mergers \citep*[e.g.][]{Naab:2009}. Finally, in any hierarchical galaxy formation scenario, dwarf galaxies are tidally disrupted and UCD or GC formation has to occur since the nuclei of dwarf galaxies are too compact to be destroyed \citep{Pfeffer:2013}. In the Milky Way, objects such as $\omega$ Cen \citep{Lee:1999,Hilker:2000} are thought to form via such a process and may be considered as `low-mass' UCDs, while the ongoing formation of a `low-mass' UCD may be observed in the M54-Sagittarius dwarf galaxy system \citep{Ibata:1997}.

Previous work predicting UCD numbers and spatial distributions from tidal stripping was performed by \citet{Bekki:2003} and \citet*{Thomas:2008}, who modelled dwarf galaxies as test particles in a static potential and used a `threshing radius' to decide if a UCD has formed or not, and \citet{Goerdt:2008}, who used the orbits of particles in a cosmological simulation combined with simulations of disc galaxies and dark matter haloes being disrupted in a static galaxy cluster. Both \citeauthor{Bekki:2003} and \citeauthor{Goerdt:2008} found their predictions matched observations, however these studies were based on a very small sample of UCDs known at the time. \citeauthor{Thomas:2008}, using a larger UCD sample, extended the analysis to lower luminosity UCDs and dwarf galaxies and found a static threshing model underpredicts UCDs at radii greater than 30 kpc for the Fornax cluster. \citet{Mieske:2012} calculated the fraction of GCs that contribute to the UCD population based on   the specific frequencies of GCs around galaxies. They found at most 50 per cent of UCDs were formed by tidal stripping.

Although they are simple to implement, static models of UCD formation in galaxy clusters have a number of disadvantages. As noted by \citet{Thomas:2008}, static models do not take into account UCD formation that may have occurred within smaller sub-clusters that later fell into the main cluster and account for the extended spatial distribution of UCDs. Galaxy clusters may have triaxial potentials, thus dwarf galaxies may be on box orbits or other chaotic orbits which provide a few close passages necessary for UCD formation but orbit at large radii at other times \citep{Pfeffer:2013}. Galaxy clusters are expected to undergo many mergers with sub-clusters during formation, which in the process may change the radial distributions of UCDs. Therefore, predictions of UCD properties within the context of cosmologically motivated galaxy cluster formation are needed to provide a definite answer on the feasibility of the tidal stripping scenario.

In this paper we use the high-resolution Millennium II cosmological simulation combined with a state-of-the-art semi-analytic galaxy formation model (described in Section \ref{sec:sims}) to predict the properties of objects formed by tidal stripping of nucleated galaxies. Throughout the paper we refer to objects formed in the simulation by tidal stripping as \textit{stripped nuclei} since such objects may resemble both GCs and UCDs and because the observed UCD populations may be the result of more than one formation channel. This paper is organized as follows. Section \ref{sec:method} describes how we identify stripped nuclei in the cosmological simulation and Section \ref{sec:results} presents the results following from our methods. In Section \ref{sec:comparison} we compare our results with observations. In Section \ref{sec:discussion} we discuss the implications of our work for UCD formation scenarios and summarize our results in Section \ref{sec:summary}.

\section{METHOD} \label{sec:method}

In this section we provide an overview of the cosmological simulations we make use of and our method for defining and identifying stripped nuclei within the simulations.

\subsection{Overview of simulations} \label{sec:sims}

Semi-analytic models of galaxy formation allow one to predict the properties of galaxies and how they evolve over cosmic time by applying analytic recipes to the dark matter merger trees of cosmological simulations. We make use of the state-of-the-art semi-analytic model of \citet[hereafter SAM]{Guo:2011} which was applied to the high-resolution Millennium-II simulation \citep[hereafter MS-II]{Boylan-Kolchin:2009}. The MS-II has a resolution 125 times that of the Millennium simulation \citep{Springel:2005}  and has a box size of 137 Mpc and a particle mass of $9.42\times 10^6 \Msun$. The SAM is constrained by low-redshift abundance and clustering in the Sloan Digital Sky Survey and is tuned to reproduce the $z=0$ mass distribution of galaxies down to stellar masses of $10^{7.5} \Msun$. For data associated with the MS-II run, we assume a cosmology consistent with the \textit{Wilkinson Microwave Anisotropy Probe} 1-year data (WMAP1) results \citep{Spergel:2003} and assume $h=0.73$ for all masses and distances. In addition, we also consider the scaled Millennium-II simulation and SAM \citep[hereafter MSII-SW7]{Guo:2013}, which was scaled to parameters consistent with a WMAP 7-year cosmology \citep[WMAP7;][]{Komatsu:2011}, to test the importance of the assumed cosmology on our results. For data associated with the MSII-SW7 run we assume parameters consistent with a WMAP7 cosmology and assume $h=0.704$ for all masses and distances. The best fit for WMAP7 is close to the best fit from Planck data \citep{Planck:2013} and therefore results would not change significantly in this cosmology.

\subsection{Simulated galaxy cluster selection} \label{sec:clusterSelection}

Since UCDs are mainly found in massive galaxy clusters \citep*[such as the Fornax and Virgo clusters which have virial masses of $7\times 10^{13} \Msun$ and $4\times 10^{14} \Msun$, respectively;][]{Drinkwater:2001, McLaughlin:1999}, we only consider SAM galaxy clusters with virial masses greater than $10^{13} \Msun/h$ ($1.37 \times 10^{13} \Msun$ in MS-II, $1.42 \times 10^{13} \Msun$ in MSII-SW7). From this selection we obtain 301 clusters in MS-II and 298 clusters in MSII-SW7. However we exclude 12 clusters in MS-II and 9 clusters in MSII-SW7 due to their proximity to the edge of the simulation box (where the edge of the box is within twice the virial radius of the cluster). We define SAM galaxy clusters at z=0 as all galaxies that are located within twice the clusters' virial radius from the central galaxy (i.e. the galaxy at the potential minimum for the cluster). To mimic observations of galaxy clusters a better choice would be to define cluster membership by the line-of-sight velocity of a galaxy. However, since we only use the SAM galaxy clusters (and their merger histories, as discussed in Section \ref{sec:mIIanalysis}) to choose the progenitors of the stripped nuclei, this makes little difference since there are typically few haloes beyond the virial radius large enough to form stripped nuclei.

\subsection{Identifying stripped nuclei} \label{sec:mIIanalysis}

\begin{figure}
  \centering
  \includegraphics[width=84mm]{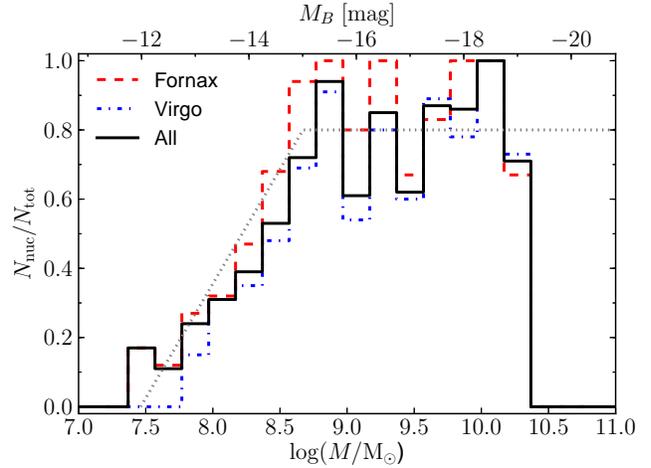}
  \caption{Fraction of early-type galaxies with nuclear clusters in the Fornax and Virgo clusters. The data sets used for the calculation are described in Section \ref{sec:mIIanalysis}. The grey, dotted line shows the adopted function for the nucleation fraction of galaxies.}
  \label{plt:fnuc}
\end{figure}

In order to identify galaxies in the simulations which may form stripped nuclei we search the galaxy merger trees of all galaxies in the SAM galaxy clusters at $z=0$. We define a galaxy as a possible stripped nucleus progenitor (hereafter referred to as candidate galaxies and the dark matter halo of the galaxies as candidate haloes) when the stellar mass first exceeds $10^{7.5} \Msun$ (i.e. all progenitors of the candidate galaxy have a stellar mass less than this limit). We choose this lower mass cut based on the nucleation fraction observed for galaxies (see Fig. \ref{plt:fnuc}) where galaxies below this mass do not host nuclei. To decide if a candidate galaxy forms a stripped nucleus we search the galaxy merger tree and find when the candidates are completely disrupted according to the galaxy disruption criteria in the SAM \citep[eq. 30 of][]{Guo:2011}. A stripped nucleus is formed in such a merger if:
\begin{enumerate}
\item The merger was a minor merger, where we define minor mergers as those with `dynamical' mass ratios smaller than 1:3 \citep[as in][]{Hopkins:2010}. The dynamical mass $M_\rmn{dyn}$ is defined as $M_\rmn{dyn} = M_* + M_\rmn{gas} + M_\rmn{DM}(<r_\rmn{s})$ where $M_*$ is the stellar mass of the galaxy, $M_\rmn{gas}$ is the cold gas mass and $M_\rmn{DM}(<r_\rmn{s})$ is the mass of the dark matter halo within the NFW scale radius. About 9 per cent of candidates have major mergers.
\item The merger happened at least 2 Gyr ago so there is enough time to form a UCD \citep{Pfeffer:2013}. About 5 per cent of candidates merged less than 2 Gyr ago. The number of objects that we remove can also give a measure of the number of recently formed objects per galaxy cluster mass (i.e. recent mergers that may be observable): for mergers within 2 Gyr we find an average of 0.65 objects per $10^{13} \Msun$ and for mergers within 1 Gyr we find an average of 0.30 per $10^{13} \Msun$. These values decrease to 0.53 per $10^{13} \Msun$ and 0.25 per $10^{13} \Msun$, respectively, when only considering nucleated objects.
\item The dynamical friction time, calculated using eq. 7-26 from \citet{Binney:1987}, is shorter than the time the stripped nucleus has been orbiting within the halo it's associated with at $z=0$. About 13 per cent of candidates will have inspiralled via dynamical friction.
\end{enumerate}
Since the particle mass ($9.42\times 10^6 \Msun$ in MS-II, $1.21\times 10^7 \Msun$ in MSII-SW7) is similar to the mass of UCDs, the most bound particle of the candidate halo in the snapshot before merging is defined as the stripped nucleus that formed after the galaxy merges. We assign the stripped nucleus a mass randomly chosen from a log-normal mass function for the nucleus-to-galaxy mass ratio with a mean of 0.3 per cent and a log-normal standard deviation of 0.5 dex, based on fig. 14 from \citet{Cote:2006}. Note we do not take into account tidal stripping for the stripped nuclei (i.e. once a stripped nucleus has formed it does not lose mass). 

The fraction of galaxies that are nucleated is taken from observations of galaxies in the Virgo and Fornax clusters. For galaxies more luminous than $M_B\leq-15$ mag the nuclear fraction was calculated from the properties of galaxies and nuclei in the ACS Fornax Cluster Survey \citep{Turner:2012} and ACS Virgo Cluster Survey (ACSVCS) \citep{Cote:2006}\footnote{We used the online version of the ACSVCS nuclei catalogue which is based on S\'ersic fits: \url{https://www.astrosci.ca/users/VCSFCS/Data_Products_files/acsvcs_nuclei_sersic.dat}}. For the fainter galaxies, $M_B>-15$ mag, we take the compilation of Fornax galaxies from \citet{Thomas:2008}. The nuclear classification is based on the Fornax Cluster Catalog \citep{Ferguson:1989}. The faint Virgo galaxies are taken from the work and classification of \citet[updated catalogue, priv. comm.]{Lisker:2007}. For the magnitude range $-15>M_B>-15.5$ the nucleation classification of the ACSVCS \citep{Cote:2006} was taken into account. We are confident that for the low surface brightness early-type dwarf galaxies the detection of nuclear clusters essentially is complete because the compact nuclei have a high contrast on top of the faint, extended stellar body of the galaxy. The fraction of nucleated galaxies according to their stellar mass is shown in Fig. \ref{plt:fnuc}. To convert from $B$-band luminosity to stellar mass, we choose a mass-to-light ratio of $(M/L)_B = 3 \ML$ for all galaxies, the average observed for dwarf galaxies in the Virgo cluster \citep{Chilingarian:2009}. We take an average nucleation fraction of 80 per cent for galaxies more massive than $M = 4.7 \times 10^8 \Msun$ ($M_B = -15$ mag). For galaxies less massive than this, we choose a fraction that varies linearly (in log-space) between 80 per cent at $M = 4.7 \times 10^8 \Msun$ and 0 per cent at $M = 3.0 \times 10^7 \Msun$ ($M_B = -12$ mag). For galaxies with masses larger than $10^{11} \Msun$ we assume nuclei no longer exist due to destruction by supermassive black holes \citep{Graham:2009}. Note that the lack of nucleated galaxies with masses $10^{10.5} < M_*/\Msun< 10^{11}$ in Fig. \ref{plt:fnuc} is due to a small sample size. We make no distinction between early- and late-type galaxies since similar nucleated fractions are observed for both \citep[e.g.][]{Carollo:1997, Boeker:2002, Cote:2006, Seth:2006}. Where possible we work with fractions of stripped nuclei instead of randomly choosing galaxies to satisfy the nucleated fraction (e.g. instead of letting only 80 per cent of disrupted galaxies form a stripped nuclei we let every galaxy create 0.8 stripped nuclei). This improves our statistics.

Two possible problems related to studying UCD formation in MS-II and MSII-SW7 are the simulation mass resolution and lack of a stellar component. The first problem relates to using a single particle to represent the nucleus of a galaxy. By using a single particle, it is possible that during a merger the particle is ejected before the merger is complete and therefore no longer accurately tracks the path of a stripped nucleus. By choosing the most bound particle of the halo just before merging as the stripped nucleus, we expect such an effect is negligible.

The second problem relates to the mass distribution of haloes in dark matter only simulations. The mass distribution of a galaxy has important implications for stripped nucleus formation since it determines how close a galaxy has to pass to the galaxy cluster centre to be disrupted (i.e. for a given mass, objects with extended mass distributions will be disrupted at larger radii than objects with more concentrated mass distributions). Haloes in dark matter only simulations are well fit by cuspy profiles \citep*[e.g. NFW profiles;][]{Navarro:1996}, while the dark matter haloes of galaxies are better fit by cored profiles \citep{Oh:2011}. The matter in a cuspy profile is more centrally concentrated compared to a cored profile, therefore making cuspy dark matter haloes harder to disrupt. However, the baryonic matter located within haloes is typically more concentrated than the dark matter and potentially acts as a compensating effect within cored haloes. It is uncertain how a combined stellar matter and cored dark matter profile compares with that of a cuspy dark matter profile and how this affects the number of stripped nuclei that are able to form in the simulations.

\begin{figure}
  \centering
  \includegraphics[width=84mm]{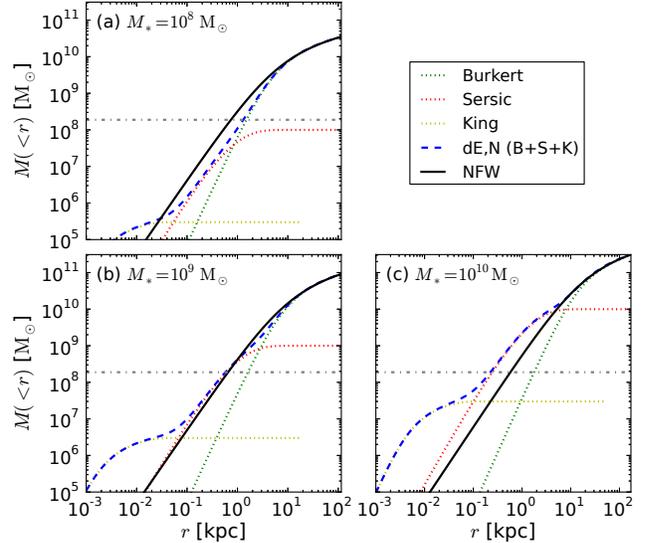}
  \caption{Cumulative mass distributions for combined King, S\'ersic and Burkert models of nucleated dwarf elliptical galaxies compared with NFW models of the same total mass from MS-II. For reference, the lowest mass halo resolved in MS-II (a halo with 20 particles) is shown by a dash-dotted line.
}
  \label{plt:massProfile}
\end{figure}

To see how differences in halo mass profiles are likely to affect our results, in Fig. \ref{plt:massProfile} we compare realistic mass profiles for nucleated dwarf elliptical galaxies (dE,Ns) with NFW profiles \citep{Navarro:1996} of the same total mass. In Fig. \ref{plt:massProfile}(b) we show the results for a dE,N with stellar mass $10^9 \Msun$ which we take to be an average dE,N. For the dwarf elliptical we chose a King profile \citep{King:1962} for the nucleus (with parameters $c=1.5$, $R_h=4$ pc, $M=3\times10^6 \Msun$), a S\'ersic profile \citep{Sersic:1963} for the stellar envelope of the galaxy (with parameters $n=1.5$, $R_e=1.0$ kpc, $M=1\times10^9 \Msun$) and a Burkert profile \citep{Burkert:1995} for the dark matter halo (with mass $M=9\times10^{10} \Msun$). The parameters for the King profile are chosen to be comparable to a typical dwarf elliptical nucleus with the mass chosen to satisfy the nucleus-to-envelope luminosity ratio of 0.3 per cent for dwarf ellipticals in the Virgo Cluster \citep{Cote:2006}. For the stellar envelope the effective radius $R_e$ is typical for a dwarf elliptical of this mass \citep{MisgeldHilker:2011} while the S\'ersic index $n$ is chosen based on the observed $M_V$-$n$ relation for early-type galaxies \citep*{Misgeld:2008}. The dark matter mass is chosen based on the average halo virial mass for galaxies in the SAM with the same stellar mass. The concentration for the NFW profile ($c_\rmn{NFW}=19.3$) is chosen using the relation for subhaloes derived by \citet*{Klypin:2011}. Remarkably, the mass profiles of the dE,N and the NFW profile in Fig. \ref{plt:massProfile}(b) match reasonably well above a mass of $10^7 \Msun$ (or a radius of 0.1 kpc), below which the nucleus dominates the mass profile of the dwarf galaxy. Therefore a dE,N with a stellar mass $10^{9} \Msun$ and a halo with an NFW profile would likely be disrupted at a similar radius in a galaxy cluster.

If we repeat the same procedure for dE,Ns of other masses the situation is slightly different. In Fig. \ref{plt:massProfile}(a) we show the comparison for a dE,N with stellar mass $10^8 \Msun$ with King profile parameters $c=1.5$, $R_h=4$ pc and $M=3\times10^5 \Msun$, S\'ersic profile parameters $n=1.5$, $R_e=0.8$ kpc and $M=1\times10^8 \Msun$ and a dark matter mass of $M=3\times10^{10} \Msun$. In Fig. \ref{plt:massProfile}(c) we show the comparison for a dE,N with stellar mass $10^{10} \Msun$ with King profile parameters $c=1.5$, $R_h=4$ pc and $M=3\times10^7 \Msun$, S\'ersic profile parameters $n=1.5$, $R_e=1.5$ kpc and $M=1\times10^{10} \Msun$ and a dark matter mass of $M=3\times10^{11} \Msun$. For a dE,N with a stellar mass $10^8 \Msun$ we find the NFW profile is more concentrated than the dE,N profile, while for a dE,N with a stellar mass $10^{10} \Msun$ we find the dE,N profile is more concentrated than the NFW profile. Therefore it is likely we underestimate formation of stripped nuclei at the low-mass end (nuclei masses less than $10^6 \Msun$) and overestimate formation of stripped nuclei at the high-mass end (nuclei masses greater than $10^7 \Msun$).

\section{RESULTS} \label{sec:results}

In this section we present the results from the analysis of the simulations described in Section \ref{sec:method}. Where not indicated otherwise, we show results from MS-II plus SAM (WMAP1 cosmology) where galaxy clusters are selected by cluster virial mass and work with fractions of stripped nuclei instead of randomly choosing galaxies to satisfy the nucleated fractions of progenitor galaxies (see Section \ref{sec:mIIanalysis}).

\begin{figure}
  \centering
  \includegraphics[width=84mm]{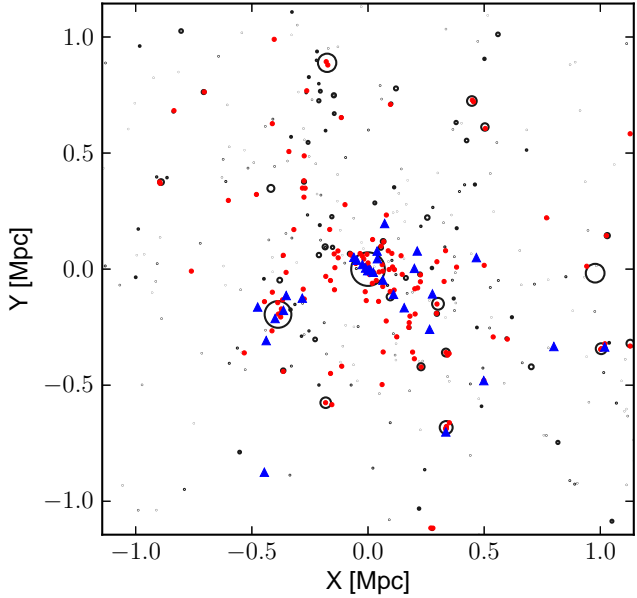}
  \caption{Projected distribution of stripped nuclei in one of the Virgo-sized galaxy clusters at $z=0$. The stripped nuclei with masses $M_\rmn{nuc} <= 10^7 \Msun$ are shown as red points and those more massive than $10^7 \Msun$ are shown as blue triangles, while galaxies are shown as black circles with radii scaling with the stellar mass. The central galaxy (the galaxy at the centre of the cluster potential) is located at (0,0) and the cluster virial radius is $r_\rmn{vir}=1.15$ Mpc, equal to the width of the box. In total there are 200 stripped nuclei in the cluster (169 within this box).}
  \label{plt:cluster}
\end{figure}

In Fig. \ref{plt:cluster} we show the projected distribution of stripped nuclei formed in a Virgo-sized galaxy cluster (for the simulated cluster the virial mass is $1.86\times 10^{14} \Msun$ and the virial radius is 1.15 Mpc). Here we randomly choose galaxies to satisfy nucleated fractions instead of working with fractions. The cluster has 200 (169) stripped nuclei in total (within the box); 103 (91) more massive than $10^6 \Msun$ and 35 (32) more massive than $10^7 \Msun$. Most of the stripped nuclei, $\sim 80$ per cent, are located within the cluster virial radius since there are few haloes beyond this radius large enough to form them. Around 60 per cent of the stripped nuclei are located within half the cluster virial radius, with most of these associated with the central galaxy. Of the stripped nuclei associated with the central galaxy, 90 per cent are located within half the cluster virial radius. These numbers are typical for most clusters.

\subsection{Number of stripped nuclei formed}

\subsubsection{Total number in clusters}

\begin{figure}
  \centering
  \includegraphics[width=84mm]{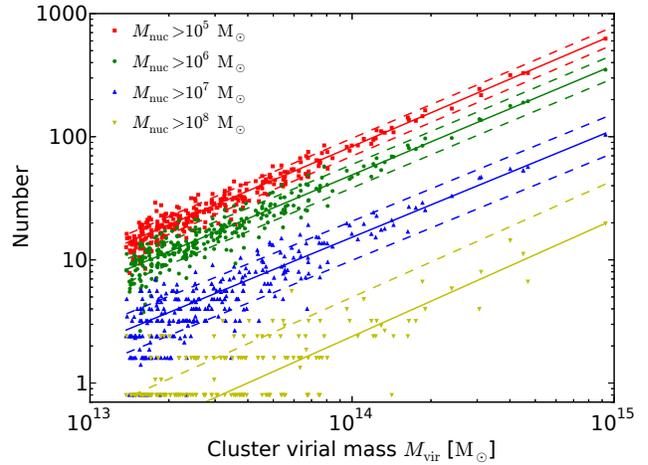}
  \caption{The number of stripped nuclei formed above a given mass located within the projected cluster virial radius at $z=0$ for individual galaxy clusters. The least-squares best fit for each population is shown by a solid line and the standard deviation of the data points from the best fit are shown by a dashed line (we don't show the error in the best fit since this is small). The equation of best fit for each population is given in the text.}
  \label{plt:absNumbers}
\end{figure}

In Fig. \ref{plt:absNumbers} we show the total number of stripped nuclei formed by $z=0$ for each galaxy cluster in the SAM compared to the cluster virial mass, along with the least-squares best fit and the standard deviation of the data points from the best fitting relation for each population. For each cluster we average the number of stripped nuclei within the projected virial radius over three sightlines (the x-, y- and z-axis of the simulation). There is very little difference between using the projected and 3D virial radius. The data shows an increase in scatter from the best fit lines for decreasing cluster masses and increasing nuclei masses which can be attributed to Poisson scatter and an increasing sample size. We find the best fitting relations of the number of stripped nuclei for a given lower mass cut to be
\begin{equation} \label{eq:numFit0}
N(M_\mathrm{nuc}>10^{5} \Msun) = 10 \pm 1.7 \, {\left(\frac{M_{\rm vir}}{10^{13} \Msun}\right)}^{0.91},
\end{equation} 
\begin{equation} \label{eq:numFit1}
N(M_\mathrm{nuc}>10^{6} \Msun) = 6.3 \pm 1.4 \, {\left(\frac{M_{\rm vir}}{10^{13} \Msun}\right)}^{0.89},
\end{equation} 
\begin{equation} \label{eq:numFit2}
N(M_\mathrm{nuc}>10^{7} \Msun) =  2 \pm 0.71 \, {\left(\frac{M_{\rm vir}}{10^{13} \Msun}\right)}^{0.87},
\end{equation} 
\begin{equation} \label{eq:numFit3}
N(M_\mathrm{nuc}>10^{8} \Msun) = 0.27 \pm 0.29 \, {\left(\frac{M_{\rm vir}}{10^{13} \Msun}\right)}^{0.95},
\end{equation} 
where $M_{\rm vir}$ is the virial mass for the cluster and the error in the relation is the standard deviation of the data points from the mean. Interestingly the slopes of the best fitting lines are slightly less than linear: formation of stripped nuclei is slightly more efficient in low-mass clusters than high-mass ones. This could be due to the fact that satellite galaxies have higher velocities in high-mass clusters, so that they are less likely to be on orbits needed for disruption. In MSII-SW7 the number of stripped nuclei predicted for clusters of similar mass is 5-10 per cent larger than in MS-II. However the slopes of the fits in MSII-SW7 are almost identical to those in MS-II.

The average number of stripped nuclei associated with satellite galaxies in clusters is 36 per cent and therefore central galaxies typically have 64 per cent of stripped nuclei. The average number associated with satellite galaxies varies with cluster mass: from 34 per cent at the low-mass end ($M_\rmn{vir} < 10^{14} \Msun$) to 58 per cent at the high-mass end ($M_\rmn{vir} > 10^{14} \Msun$). 

On average 22 per cent of stripped nuclei were formed in haloes which then merged into another halo. When looking only at the central galaxies in galaxy clusters, this jumps to 29 per cent on average; 28 per cent for clusters with virial masses $M_\rmn{vir} < 10^{14} \Msun$ and 35 per cent for clusters with virial masses $M_\rmn{vir} > 10^{14} \Msun$.

\subsubsection{Numbers in individual galaxies} \label{sec:individualNumbers}

\begin{figure*}
\centering
  \includegraphics[width=0.49\textwidth]{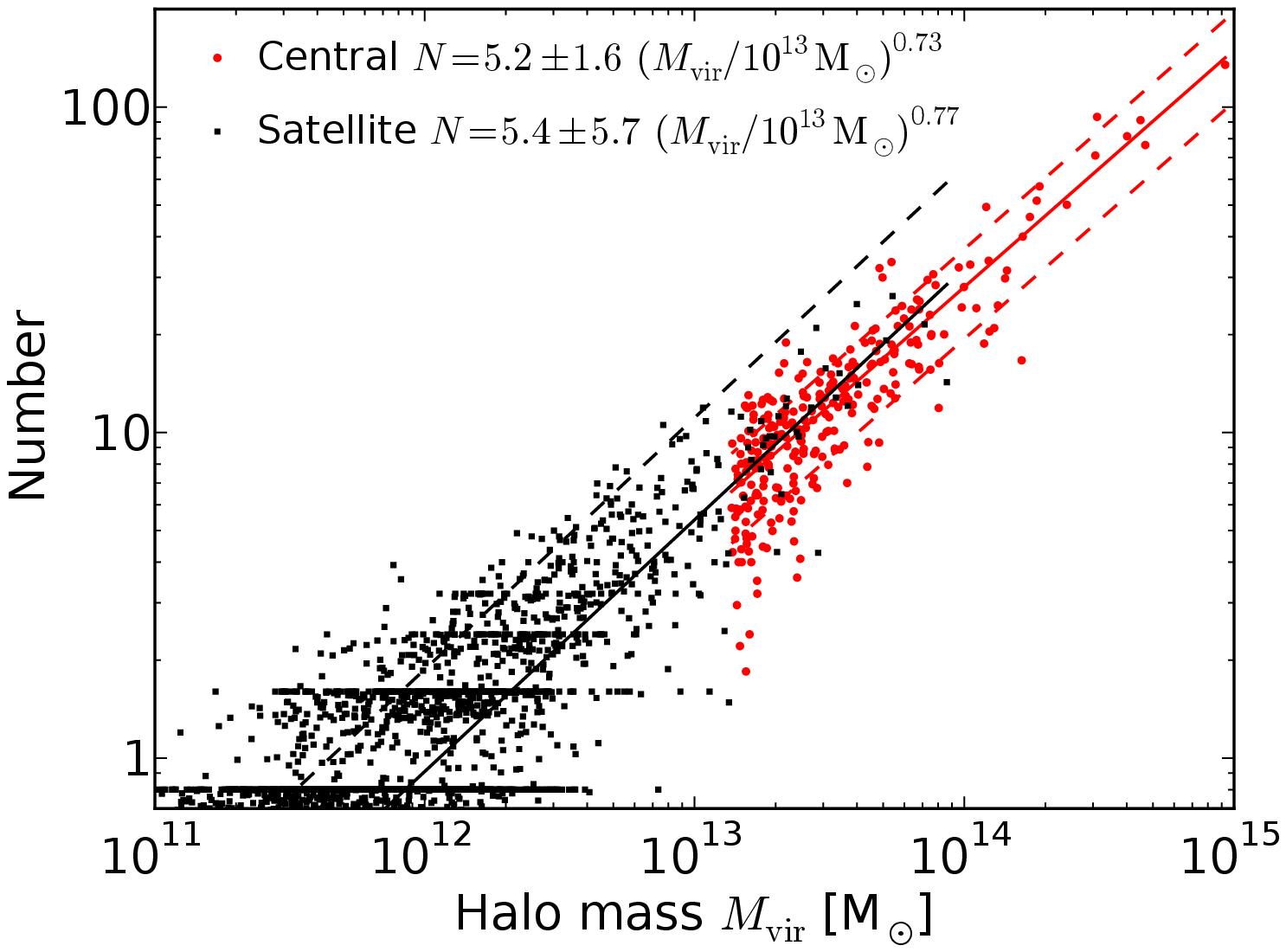}
  \includegraphics[width=0.49\textwidth]{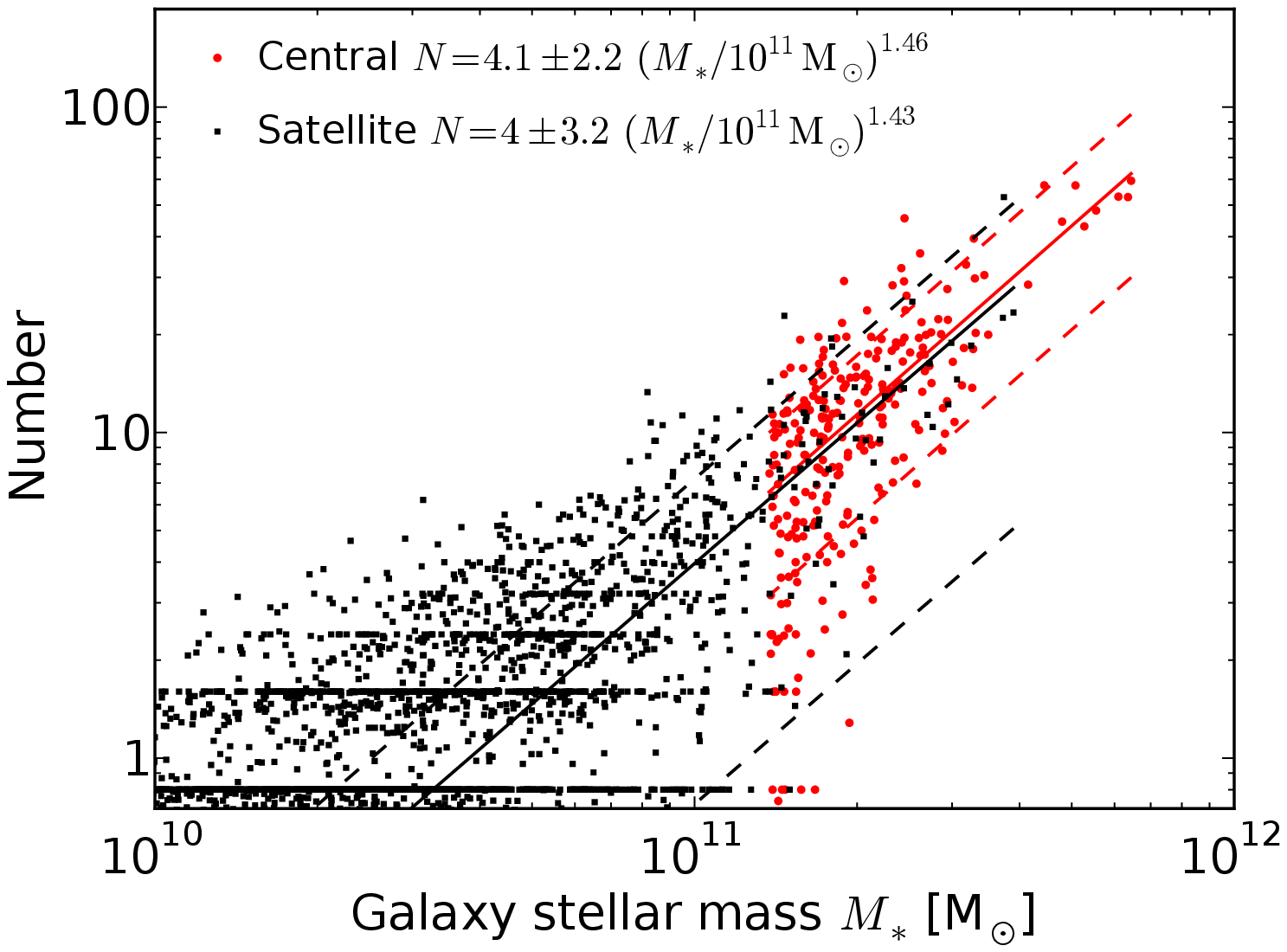}
  \caption{Number of stripped nuclei with masses larger than $10^6 \Msun$ formed for individual galaxies in the simulated galaxy clusters compared to the halo virial mass of the galaxy (left) and galaxy stellar mass from the semi-analytic model (right). Central galaxies (galaxies at the centre of a galaxy cluster) are shown as red dots, while satellite galaxies (all other galaxies) are shown as black squares. For satellite halo masses we take the virial mass before infall into a cluster. The least-squares best fit for each population is shown by a solid line and the standard deviation of the data points from the best fit line is shown by a dashed line (we don't show the error in the best fit since this is small). The equation of best fit for each population is shown in the legend of the figures. In the right panel we select galaxy clusters in the SAM by central galaxy mass (described in Section \ref{sec:individualNumbers}), rather than cluster virial mass, so that the number of central galaxies is complete above $10^{11} \Msun/h$.}
  \label{plt:individualNumbers}
\end{figure*}

In this section we compare the number of stripped nuclei that form around individual galaxies. Specifically, we compare the numbers around central and satellite galaxies in the galaxy clusters and compare how the numbers correlate with the halo virial mass and the stellar mass of the galaxy. To determine which galaxy and halo each stripped nucleus is associated with, we trace the galaxy and halo merger trees of the candidate galaxies and haloes to the descendants at $z=0$. We assume that when a galaxy or halo becomes disrupted by its host galaxy or halo, any stripped nuclei from the satellite are transferred to the host. Before a satellite is disrupted we assume that all stripped nuclei it hosted before infall are still associated with it, even though some nuclei may have been unbound during tidal stripping. For all satellite halo masses in this section we take the virial mass before infall into the host halo (and subsequent tidal stripping). This allows us to directly compare the number of stripped nuclei for central and satellite galaxies.

As we selected the sample of galaxy clusters from the SAM by virial mass only (see Section \ref{sec:clusterSelection}), our original sample of clusters only has a complete sample of central galaxies above stellar masses of $\sim 3\times 10^{11} \Msun$. Therefore, in order to obtain a complete sample to lower masses we create a new sample of clusters which are selected to have a central galaxy stellar mass larger than $10^{11} \Msun/h$. This gives us a sample of 271 clusters which we analyse using the method described in Section \ref{sec:mIIanalysis}. The new sample of clusters is therefore used for the right panel in Fig. \ref{plt:individualNumbers}, while the original sample is used for the left panel, where both panels are a WMAP1 cosmology.

In Fig. \ref{plt:individualNumbers} we compare the number of stripped nuclei with masses larger than $10^6 \Msun$ that form for each galaxy against the halo virial mass and stellar mass of the galaxy. The left panel of the Fig. \ref{plt:individualNumbers} shows that the number of stripped nuclei which form for both central and satellite haloes follows a tight relation with halo mass. The best fitting relation for each sample is almost identical, which suggests they are part of the same distribution and it is not necessary to make a distinction between central and satellite galaxies when comparing against halo mass. As in Fig. \ref{plt:absNumbers} most of the points in the figure are within a factor of two from the best fitting relation, although there is slightly larger scatter at the low-mass end. Compared to Fig. \ref{plt:absNumbers}, the best fitting relation for the number of stripped nuclei with halo mass is slightly flatter (in Fig. \ref{plt:absNumbers} the power-law slope is $\alpha \approx 0.9$, compared to $\alpha \approx 0.75$ in Fig. \ref{plt:individualNumbers}). This can be attributed to more massive haloes having a higher proportion of satellites that can host stripped nuclei compared to lower mass haloes. In the right panel of Fig. \ref{plt:individualNumbers}, as in the left panel, the relations for centrals and satellites are almost identical, suggesting they are part of the same distribution. However there is a much larger scatter in the number of stripped nuclei for a given stellar mass of a galaxy compared to the left panel. We therefore suggest that the halo mass of a galaxy is a better predictor for the number of stripped nuclei formed by a galaxy than the galaxy stellar mass.

\subsubsection{Most massive stripped nuclei formed} \label{sec:maxMass}

\begin{figure*}
  \centering
  \includegraphics[width=0.49\textwidth]{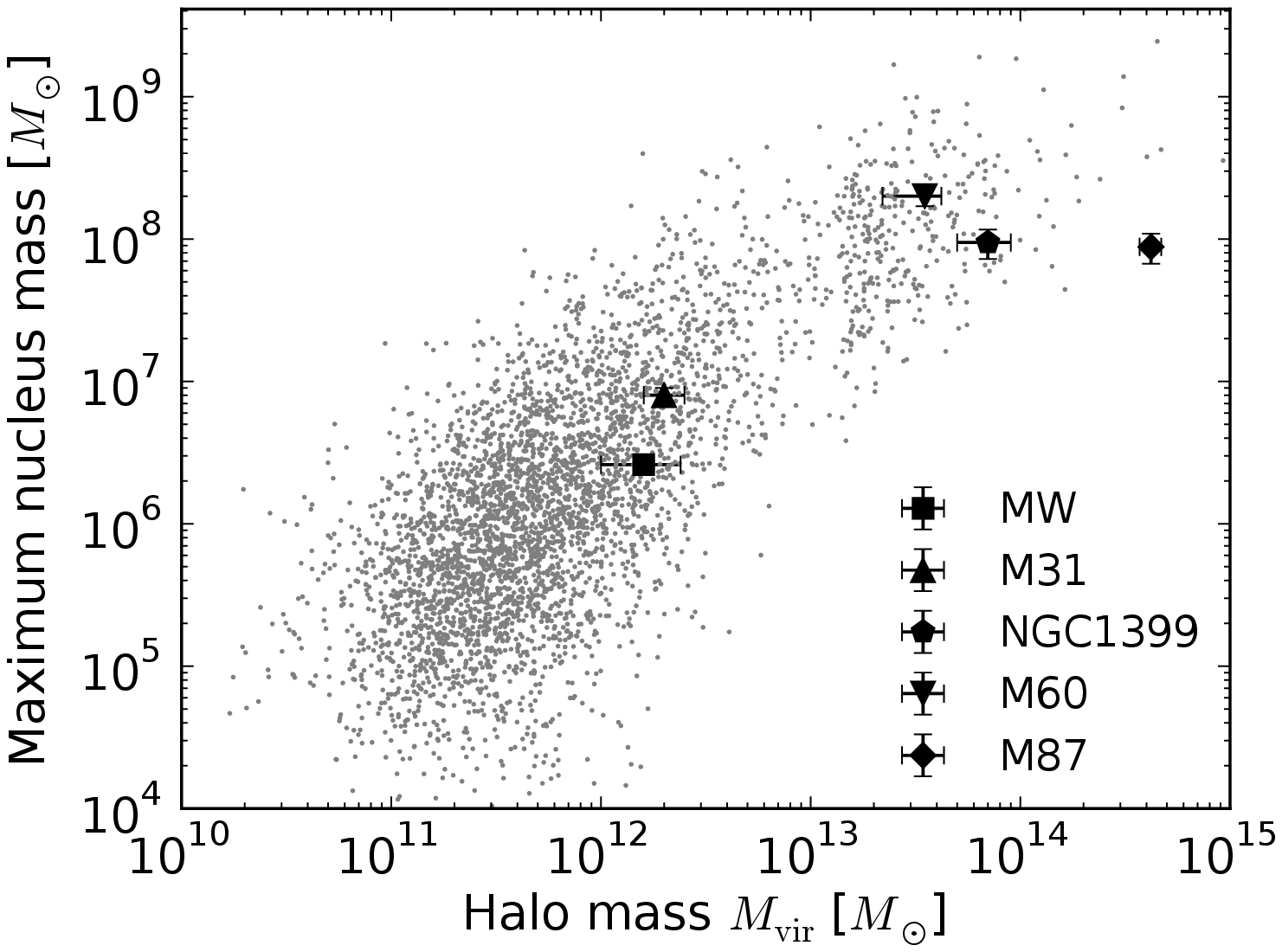}
  \includegraphics[width=0.49\textwidth]{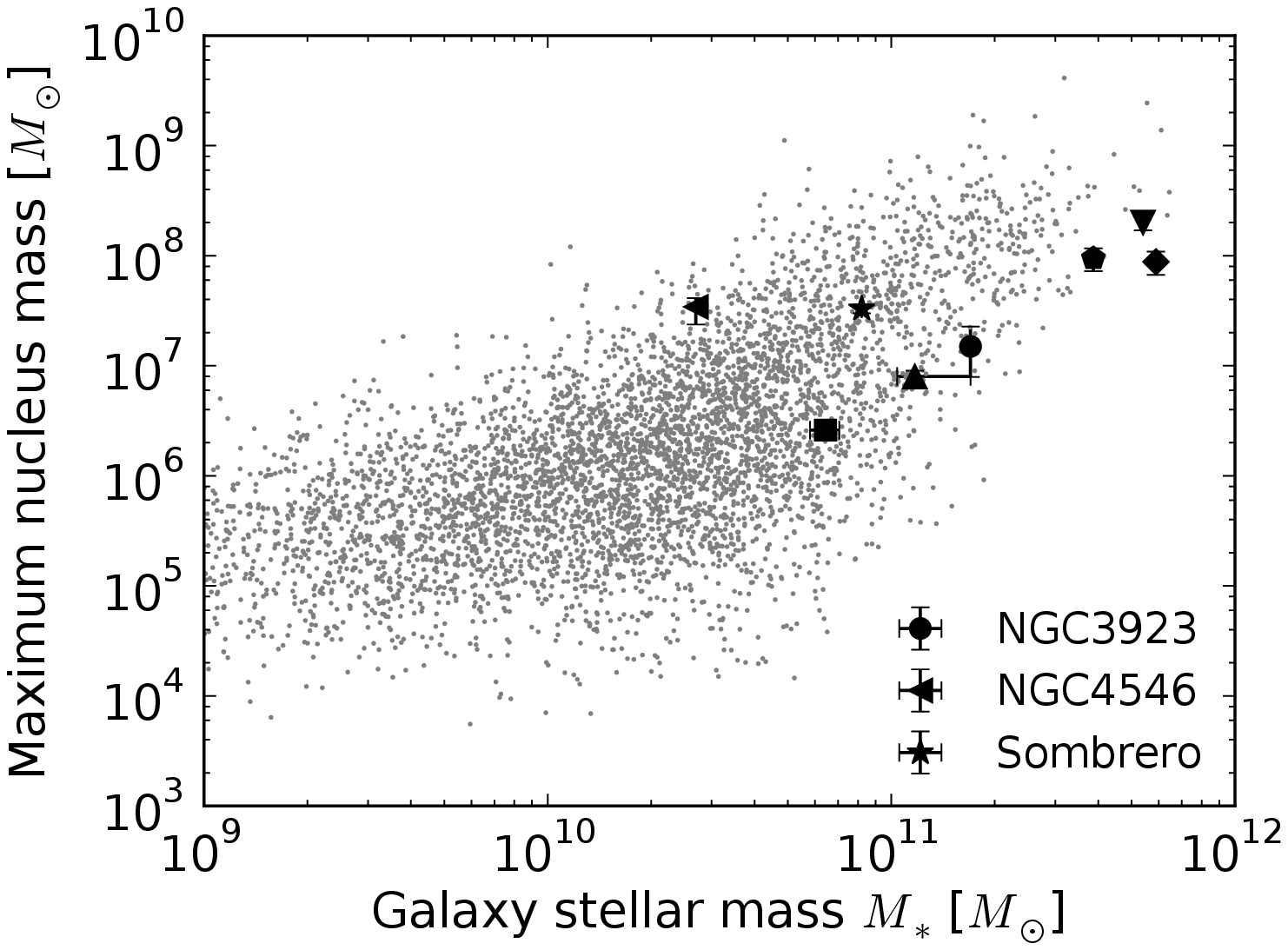}
  \caption{Maximum stripped nucleus mass for individual galaxies compared to the halo virial mass or halo mass before infall for satellite haloes (left) or galaxy stellar mass (right) of the galaxy. For the Milky Way, M31, NGC 1399, M87 and M60 the most massive GC or UCD ($\omega$ Cen, G1, UCD3, VUCD7 and M60-UCD1, respectively) are shown. The masses of $\omega$ Cen, G1, UCD3, VUCD7 and M60-UCD1 are taken from \citet{Jalali:2012}, \citet{Baumgardt:2003}, \citet{Hilker:2007}, \citet{Evstigneeva:2007b} and \citet{Strader:2013}, respectively. The halo masses of the Milky Way, M31, NGC 1399, M87 and M60 are taken from \citet{Boylan-Kolchin:2013}, \citet{Fardal:2013}, \citet{Drinkwater:2001}, \citet{McLaughlin:1999} and \citet{Humphrey:2006}, respectively, where we assume NGC 1399 and M87 sit at the centre of the cluster potential. The stellar masses of the Milky Way and M31 are taken from \citet{McMillan:2011} and \citet{Geehan:2006}, respectively, and NGC 1399, M87 and M60 from \citet{MisgeldHilker:2011}. In the right panel we also include the most massive UCDs of NGC 3923, NGC 4546 and the Sombrero galaxy from \citet{Norris:2011}.}
  \label{plt:maxMass}
\end{figure*}

Given that galaxies and haloes of larger masses are able to tidally strip satellites of larger mass without undergoing a major merger, it is expected that the most massive stripped nucleus for a halo should scale with halo mass. In the left panel of Fig. \ref{plt:maxMass} we compare the mass of the most massive stripped nucleus formed for each galaxy with the halo virial mass of the galaxy. We do not distinguish between central and satellite galaxies since we showed in Section \ref{sec:individualNumbers} it is unnecessary when comparing against halo mass (halo mass before infall for satellite galaxies). Here we randomly choose galaxies to satisfy nucleated fractions instead of working with fractions. In the right panel of Fig. \ref{plt:maxMass} we compare the most massive stripped nucleus of a galaxy with the galaxy stellar mass. Again we do not distinguish between central and satellite galaxies. The maximum stripped nucleus mass is largely set by the major merger prescription and therefore scales with halo and galaxy mass. The large scatter in the maximum nucleus mass for a given halo or stellar mass can be attributed to the particular merger histories of each galaxy, as well as the distribution in the nucleus-to-galaxy mass ratio.

A tidal stripping origin has been suggested for the most massive GCs in the Milky Way and M31 and is also the most likely origin for the most massive UCDs in the Virgo and Fornax clusters. The most massive GCs in the Milky Way and M31 are $\omega$ Cen and G1, respectively. For the Virgo and Fornax clusters the most massive UCDs are M60-UCD1 and UCD3, respectively. These are shown in Fig. \ref{plt:maxMass} and agree well with the predicted maximum nucleus masses for halo mass, but lie slightly under the prediction for stellar mass. We also show the most massive UCDs from NGC 3923, NGC 4546 and the Sombrero galaxy in the right panel of Fig. \ref{plt:maxMass}. NGC 4546-UCD1 falls above the most massive nucleus we predict which may partially be caused by recent star formation \citep{Norris:2011}. A tidal stripping origin for these GCs and UCDs is therefore compatible with our results.

With the exception of the Virgo and Fornax clusters and NGC 4546, the masses of the most massive GCs and UCDs also agree well with that predicted from the GC luminosity functions \citep{Hilker:2009, Norris:2011}. Therefore an agreement between the predicted maximum stripped nucleus mass and that observed for a galaxy does not necessarily imply a tidal stripping origin.

\subsection{Mass function of stripped nuclei} \label{sec:massFunction}

\begin{figure}
  \centering
  \includegraphics[width=84mm]{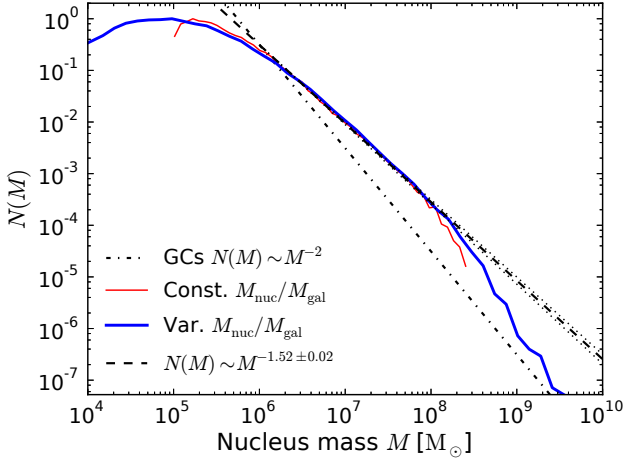}
  \caption{Normalized mass function of the stripped nuclei for all clusters with a constant nucleus mass fraction $M_\mathrm{nuc}/M_\mathrm{gal}=0.3$ per cent (thin red solid line) and distribution in the mass fraction (thick blue solid line) compared with the mass function of GCs (dash-dotted line; arbitrarily scaled such that GCs have the same absolute number at $10^6 \Msun$). The best-fitting slope for the stripped nuclei with masses between $10^6$ and $10^8 \Msun$ for a constant nucleus mass fraction is shown by the dashed line, with the standard deviation in the slope shown by the dotted lines.}
  \label{plt:massFunction}
\end{figure}

In Fig. \ref{plt:massFunction} we show the predicted mass function for the stripped nuclei from all clusters in MS-II compared to the mass function of GCs \citep[a power-law with a slope $\alpha \simeq -2$ for GCs more massive than $3\times 10^5 \Msun$,][]{Jordan:2007}. Between masses $10^6 < M/\Msun < 10^8$ the nuclei follow a power-law with a slope $\alpha = -1.52 \pm 0.02$. We find no systematic variation of the slope of the mass function with galaxy cluster mass. We also compared this result to the prediction from the MSII-SW7 run and found the result is unchanged. For low-mass galaxy clusters ($M_\rmn{vir} \sim 10^{13}\Msun$) the slope can vary significantly due to low numbers of nuclei and Poisson scatter, however high-mass clusters ($M_\rmn{vir} > 10^{14}\Msun$) have little scatter from the average. Below masses of $10^6 \Msun$ the mass function of the nuclei flattens due to the decreasing nucleation fraction of the progenitor galaxies. Above masses of $10^8 \Msun$ the nuclei mass function steepens due to the steepening of the mass function of the progenitor galaxies with stellar masses above $10^{10.5}\Msun$ \citep[see fig. 7 of][]{Guo:2011}. Since stripped nuclei have a flatter mass function than GCs, the contribution of stripped nuclei to UCDs will be more important at the high-mass end.

We note that this result doesn't take into account that stripped nuclei may retain some stars from the main galaxy \citep{Pfeffer:2013}. If the average difference in size between UCDs and the nuclei of the progenitor galaxies is a factor of two \citep{Evstigneeva:2008}, this will increase the mass by 50 per cent and therefore will not be significantly different from our prediction \citep[assuming the increase in size is due to stripped nuclei retaining some mass from the progenitor galaxy;][]{Pfeffer:2013}. In addition, we do not include ongoing tidal stripping of the objects, which would act in the opposite direction. Assuming the efficiency of these processes does not depend on the mass of the progenitor galaxy the slope of the mass function would change little, although how they affect the absolute scale is unclear.

\subsection{Ages of disrupted galaxies}

\begin{figure}
  \centering
  \includegraphics[width=84mm]{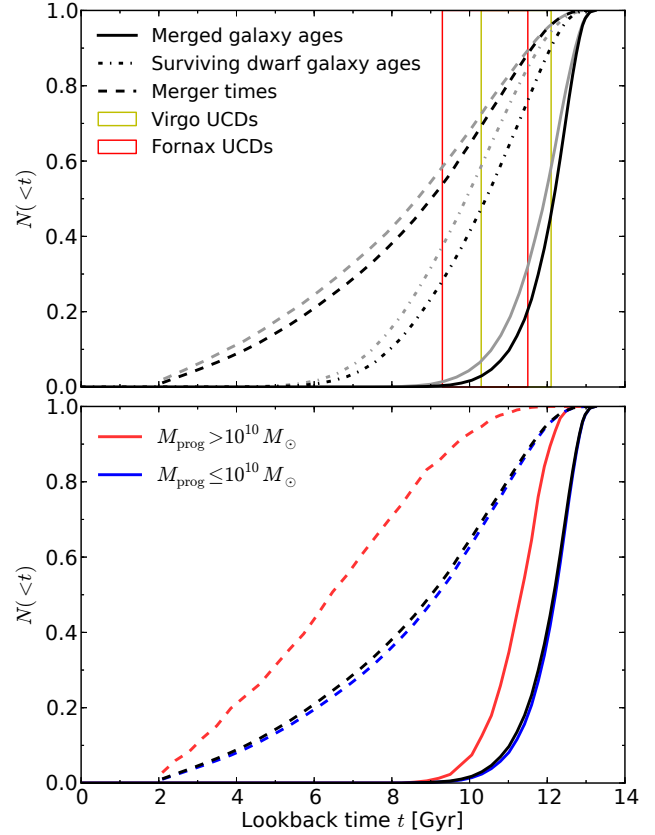}
  \caption{Normalized cumulative distribution of the predicted ages and merger times for disrupted galaxies. The \textit{merged galaxy ages} are the mass-weighted ages of the galaxies from the SAM which are disrupted to form stripped nuclei in our model. The \textit{surviving dwarf ages} are the mass-weighted ages of the dwarf galaxies (with masses between $10^{7.5}$-$10^{10.5} \Msun$) which survive in the SAM galaxy clusters at $z=0$. The \textit{merger time} shows the time when the dark matter haloes of the progenitor galaxies which form stripped nuclei are no longer resolved. In the upper panel black lines show the results from a WMAP1 cosmology, while grey lines show results from a WMAP7 cosmology. We also show the standard deviation of ages of UCDs in the Virgo and Fornax clusters for comparison \citep{Francis:2012}. In the lower panel we divide the progenitor galaxies into high-mass and low-mass groups and compare the galaxy ages and merger times for a WMAP1 cosmology. Since the low-mass group contains most of the galaxies its distributions are almost identical to that of the total population.}
  \label{plt:mergerTimes}
\end{figure}

In the upper panel of Fig. \ref{plt:mergerTimes} we show the predicted mass-weighted ages and merger times for the galaxies which are disrupted to form stripped nuclei, as well as the mass-weighted ages of dwarf galaxies which survive to $z=0$. The figure shows that galaxies which merge and become stripped nuclei mostly form very early (95 per cent formed more than 10.5 Gyr ago), even though mergers happen up until 2 Gyr ago (the minimum time we require to form a stripped nucleus). The dwarf galaxies remaining at $z=0$, however, are typically younger by about 2 Gyr than the dwarf galaxies that merge. This preferential disruption of old galaxies agrees well with the observation that typically only old dwarf galaxies are located near the centre of galaxy clusters \citep*[e.g.][]{Paudel:2011}. The plot also shows that there is very little difference between WMAP1 and WMAP7 cosmologies for the ages of galaxies which merge to form stripped nuclei and the merger times of disrupted galaxies. However the surviving dwarf galaxies are slightly younger ($\sim 0.5$ Gyr) in a WMAP7 cosmology.

From the ages of the progenitor galaxies alone it is impossible to infer ages for the stripped nuclei since the ages of nuclei bear little relation to their host galaxies \citep{Paudel:2011}. For our scenario we require that some part of the nucleus is built up with the formation of the galaxy or very soon afterwards (within $0.9$ Gyr) in order to explain 95 per cent of the stripped nuclei we predict. This would particularly affect the most massive nuclei since they may take a longer time to form than low-mass nuclei if they are built up by several star formation events. Therefore, in the lower panel of Fig. \ref{plt:mergerTimes} we divide the galaxies into high-mass (galaxy stellar mass $M > 10^{10} \Msun$) and low-mass (galaxy stellar mass $M \leq 10^{10} \Msun$) groups. The figure shows that high-mass galaxies typically form about 0.8 Gyr later than low-mass galaxies, however, they also merge much later than low-mass galaxies ($\sim 2.5$ Gyr later). To explain 95 per cent of the high-mass stripped nuclei (masses $M \gtrsim 10^7 \Msun$) it is only required that the nuclei form within $\sim$2 Gyr.

In the upper panel of Fig. \ref{plt:mergerTimes} we also plot the mean ages of UCDs in the Virgo and Fornax clusters from \citet{Francis:2012} for comparison. If nucleus formation happens (or continues to happen) some time between the formation and the merging of the progenitor galaxy, the ages of stripped nuclei would fall well within the measured ages of UCDs.

\subsection{Radial distribution of stripped nuclei}

\begin{figure*}
\centering
  \includegraphics[width=\textwidth]{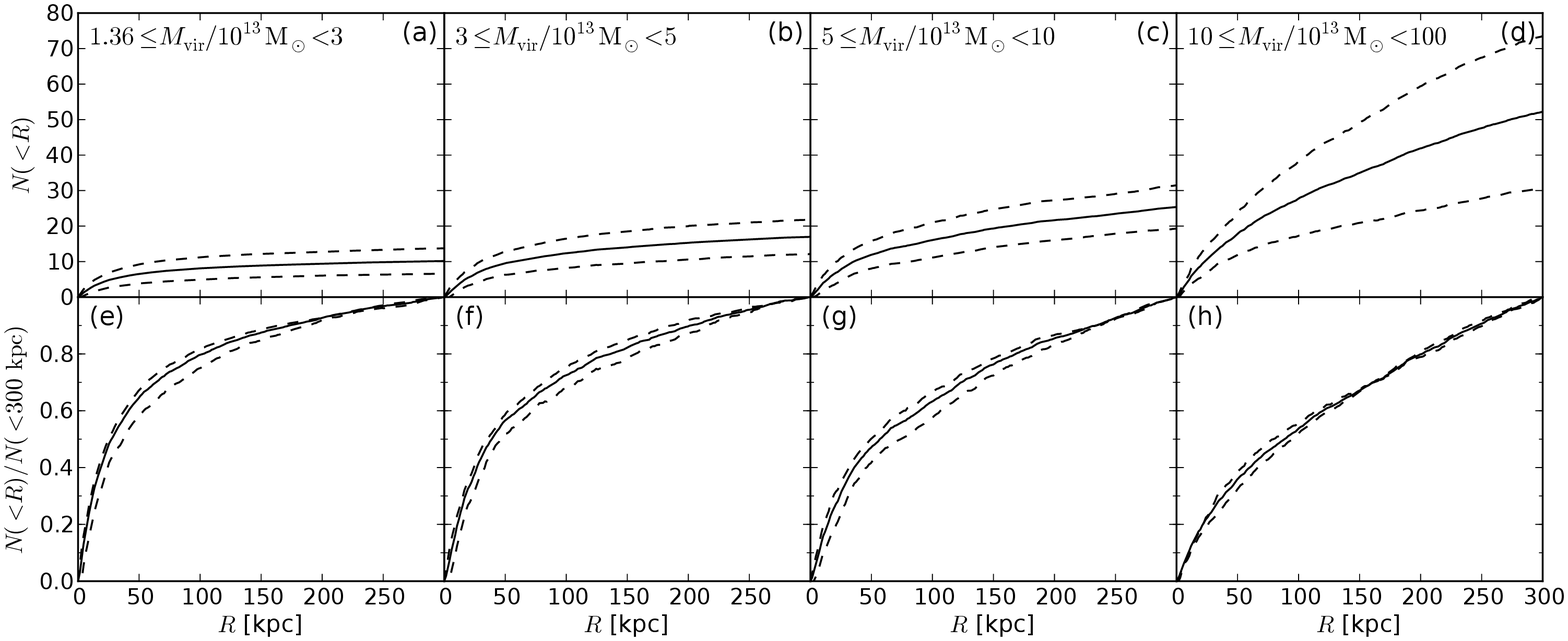}
  \includegraphics[width=\textwidth]{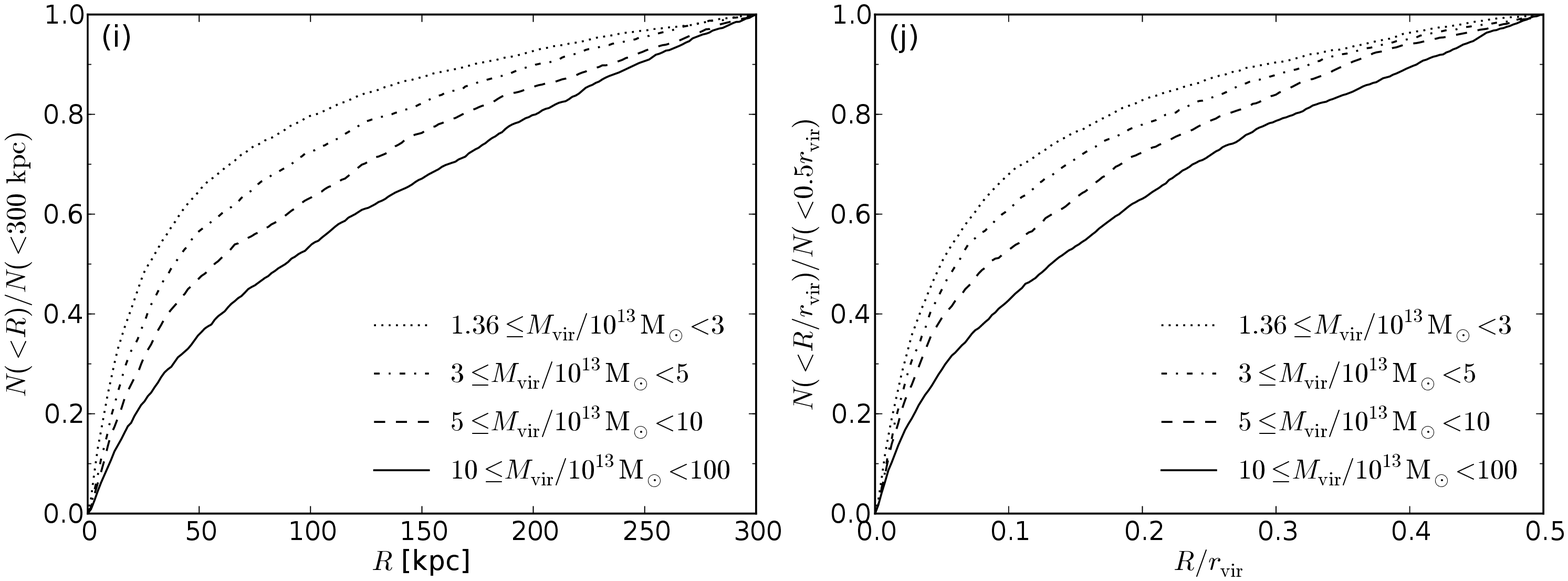}
  \caption{Projected cumulative radial distributions from the centre of the galaxy cluster for stripped nuclei with masses larger than $10^6 \Msun$. The panels (a)-(d) show the radial distributions within 300 kpc with the data stacked by the virial mass of the cluster (the mass ranges are indicated in each panel). The solid line shows the mean for each sample, while the dashed lines show the standard deviation. Panels (e)-(h) show the normalized distributions of panels (a)-(d). The mean and standard deviation for the absolute and normalized distributions were calculated separately. Panels (i) and (j) show the mean of the normalized radial distributions for each mass range. Panel (i) shows the radial distributions within 300 kpc, while panel (j) shows projected radius scaled by the virial radius for each cluster.}
  \label{plt:radialDist}
\end{figure*}

In Fig. \ref{plt:radialDist} we show the projected cumulative radial distributions from the centre of a galaxy cluster for the stripped nuclei more massive than $10^6 \Msun$, where the clusters are binned according to cluster mass. The number of simulated galaxy clusters in each mass range are 170, 58, 40 and 21, from least massive bin to most massive. For each cluster we average over three sightlines (the x-, y- and z-axis of the simulation) to calculate the radial distribution. Note that the radial distributions include stripped nuclei associated with satellite galaxies and not just those associated with the central galaxy. From least massive bin to most massive in panels (a)-(i) the percentage of stripped nuclei associated with satellite galaxies is approximately 18, 20, 25 and 28 per cent and in panel (j) it is approximately 17, 22, 30 and 43 per cent.

Panels (a)-(d) show the absolute numbers of stripped nuclei while panels (e)-(h) show the normalized radial distributions. Despite the large differences in the absolute numbers (the standard deviation for all mass bins is $\sim 25$ per cent of the mean) the normalized radial distributions have a small deviation from the mean (less than 5 per cent). In panel (i) and (j) we compare the normalized radial distributions for each mass bin. The radial profiles strongly depend on the cluster mass, with the radial profiles being more concentrated in low-mass clusters. This is partially explained by high-mass clusters having more stripped nuclei associated with satellite galaxies in the cluster, and therefore a larger number of stripped nuclei at larger projected distances. If we only plot the nuclei associated with the central galaxy panel (i) is almost unchanged, while in panel (j) the spread between the largest and smallest mass bins is reduced by about 50 per cent. The remaining difference can be explained by high-mass clusters having more mass within a given radius (or fraction of the virial radius) than low-mass clusters and can therefore tidally strip galaxies at a larger radius.

Comparing these results with observations is not trivial since one needs uniform observations that cover a significant fraction of a galaxy cluster. Such observations might become available with programs such as the Next Generation Virgo cluster Survey \citep[NGVS;][]{Ferrarese:2012}.

\subsection{Orbits of stripped nuclei} \label{sec:orbits}

One question that remains unanswered from \citet{Pfeffer:2013} is whether the orbits that can explain the full size range of UCDs in galaxy clusters are likely to occur. \citeauthor{Pfeffer:2013} found that dwarf galaxies should have at most a few close passages within $\sim 10$ kpc of the central galaxy in a rich galaxy cluster to form extended UCDs (those that are many times the size of dwarf galaxy nuclei of a similar mass) while those with many close passages form objects approximately the size of the nucleus. The orbits that progenitor galaxies take in galaxy clusters during the formation of GCs/UCDs therefore have strong implications for the size of objects that can form during tidal stripping.

The snapshot output times of MS-II (every 300 Myr on average between 0 and 12 Gyr in lookback time) is much too coarse to follow the orbits of galaxies which are likely to become stripped nuclei. Therefore in order to find when peri-/apocentres occur we advance the particles between snapshots based on their current position and velocity. For each candidate we find the snapshot at which the candidate halo first becomes a satellite in a larger halo, then at each snapshot until z=0 we find the position and velocity of candidate halo or the most bound particle which we designated the stripped nucleus (once the candidate halo is disrupted). We use a simple leapfrog method to advance the halo or particle between the snapshots with 1000 time steps. We take into account the gravitational potential of all subhaloes in the clusters for the potential calculation and assume an NFW profile for all haloes. The positions of all haloes are also advanced between snapshots. When determining the number of pericentre passages we assume only passages with the host halo are important for disrupting the galaxies. Since we don't know how the mass of the subhaloes change between the snapshots, we advanced the halo or particle at one timestep forward, then at the next timestep backward, and chose the orbit that fits the positions of the particles in the snapshots.

Since this is a rather simple and na\"ive analysis, there are a number of problems which may introduce significant errors. The two problems which will introduce the largest errors in the orbit calculations are not taking into account dynamical friction and not taking into account any triaxiality of haloes. Since the halo catalogues of MS-II do not include any information on triaxiality it is not possible to include this effect in this work. We assume here that all pericentre passages are of equal importance to the formation of the stripped nuclei. However, since some passages may have occurred in smaller haloes before entering a galaxy cluster, we may overestimate the number of important pericentre passages to the stripped nuclei. Therefore the number of extended stripped nuclei may be larger than we predict. For these reasons we do not use the orbits to decide whether a stripped nucleus has formed and instead rely on the merger tree. The following results are therefore only approximate.

\begin{table}
\centering
\caption{Mean and median number of pericentre passages and fraction of objects with at least one pericentre passage less than a given distance.}
\label{tab:passages}
\begin{tabular} {@{}lccc@{}}
  \hline
  Pericentre & Mean & Median & Fraction (\%) \\
  \hline
  $< 5$ kpc & 23 &  9 & 81 \\
  $<10$ kpc & 26 & 12 & 86 \\
  $<20$ kpc & 28 & 15 & 91 \\
  $<50$ kpc & 31 & 19 & 96 \\
  \hline
\end{tabular}
\end{table}

In Table \ref{tab:passages} we show the typical number of pericentre passages for stripped nuclei with at least one passage less than a given distance. We find most candidates, 66 per cent, typically have more than three pericentre passages less than 10 kpc and therefore likely form compact objects similar to the size of the initial nucleus \citep[within a factor of $\sim 2$,][]{Pfeffer:2013}. We also find 20 per cent of candidates have only between one and three pericentre passages less than 10 kpc and therefore satisfy the condition to form extended UCDs. Alternatively, candidates that have no pericentre passages less than 20 kpc may also form extended objects and which account for 9 per cent of candidates. However, these are mainly dwarf galaxies with stellar masses between $10^{7.5}$ and $10^9 \Msun$ (95 per cent of galaxies with stellar masses greater than $10^{10}\Msun$ have a least one orbit less than 20 kpc) and therefore will most likely still form compact objects since they will be disrupted at larger distances than more massive galaxies. This supports the view that there will be few intermediate objects between UCDs and dwarf galaxies since objects that form slowly are rare \citep[e.g.][]{Brodie:2011, Bruens:2012}. The final 5 per cent of candidates have no passages less than 10 kpc and at least one between 10 and 20 kpc. Since almost all of these galaxies have have low masses (less than 5 per cent have masses larger than $10^{10}\Msun$) they will also likely form compact objects.

\section{COMPARISON WITH OBSERVATIONS} \label{sec:comparison}

\subsection{Milky Way} \label{sec:MW}

In the Milky Way it has been suggested that the GCs $\omega$ Cen \citep{Lee:1999,Hilker:2000}, M22 \citep{Marino:2009}, NGC 1851 \citep{Han:2009}, Terzan 5 \citep{Ferraro:2009}, NGC 2419 \citep{Cohen:2010}, NGC 3201 \citep{Simmerer:2013} and NGC 5824 \citep{Saviane:2012} are the nuclei of disrupted dwarf galaxies due to either stellar populations with multiple ages or stellar populations with a spread in heavy element abundances. We use the GC luminosities from the online GC database of \citet{Harris:1996} and the mass-to-light ratios from \citet{McLaughlin:2005}, from which we use the Wilson fits since the fits are at least as good, and often better, than the King and power-law fits \citep[with the exception of Terzan 5 where we take the mass from][]{Lanzoni:2010}. This gives seven GCs with masses larger than $10^5 \Msun$ (all suggested GCs) and two GCs more massive than $10^6 \Msun$ ($\omega$ Cen and Terzan 5) that are thought to be remnants of stripped dwarf galaxies. If we extrapolate equations \ref{eq:numFit0}-\ref{eq:numFit3} to the virial mass of the Milky Way $M_\rmn{vir} = 1.6_{-0.6}^{+0.8}  \times 10^{12} \Msun$ \citep[90 per cent confidence interval from][]{Boylan-Kolchin:2013} we predict $1.9^{+1.3}_{-0.9}$ stripped nuclei have formed around the Milky Way with a mass larger than $10^5 \Msun$ and $1.2^{+0.9}_{-0.6}$ have formed with a mass larger than $10^6 \Msun$. Around $41^{+38}_{-23}$ per cent of Milky Way mass haloes are predicted to have a stripped nucleus more massive than $10^7 \Msun$ and $5^{+10}_{-5}$ per cent are predicted to have a stripped nucleus more massive than $10^8 \Msun$. The number of stripped nuclei predicted for the Milky Way in MSII-SW7 are very similar to that in MS-II. Therefore, given the errors from the fits in Fig. \ref{plt:absNumbers} and the virial mass estimates, the number of stripped nuclei that we predict with masses larger than $10^6 \Msun$ is completely consistent with the number that are observed in the Milky Way. For stripped nuclei with masses larger than $10^5 \Msun$ the number we predict is a factor of three lower than the number that are observed in the Milky Way. This does not necessarily imply tension with observations since the Poisson scatter is large for Milky Way-sized haloes and it is not clear whether these GCs in the Milky Way are definitely stripped nuclei or have formed via a different process. Alternatively this disagreement could be explained by low-mass dwarf galaxies having a higher nucleation fraction at early times.

\subsection{M31}

Four GCs around M31 have also been suggested to be the nuclei of disrupted dwarf galaxies: G1 \citep{Meylan:2001} and G78, G213 and G280 \citep{Fuentes-Carrera:2008}. These GCs all have masses larger than $10^6 \Msun$ \citep*{Strader:2011}. Current estimates suggest a halo virial mass for M31 close to the upper bound for the Milky Way \citep[$M_\rmn{vir} = 2_{-0.4}^{+0.5} \times 10^{12} \Msun$][]{Fardal:2013}. Therefore we expect M31 to host $1.5^{+0.7}_{-0.5}$ stripped nuclei with masses larger than $10^6 \Msun$, a factor of two lower than the number suggested by observations. However in Fig. \ref{plt:individualNumbers} some haloes with masses of $\sim 2\times 10^{12} \Msun$ do host four stripped nuclei with masses larger than $10^6 \Msun$ and therefore M31 may have had more mergers than typical for it's halo mass.

\subsection{Fornax cluster} \label{sec:Fornax}

\begin{table}
\centering
\caption{Number of predicted stripped nuclei and the number of confirmed UCDs observed in the Fornax cluster within a given projected radius. The observed number of UCDs with masses $M>2\times10^6 \Msun$ represent lower limits because of incompleteness.}
\label{tab:FornaxNumber}
\begin{tabular} {@{}ccccc@{}}
  \hline
  Mass ($\Msun$) & \multicolumn{2}{c}{$R<83$ kpc} & \multicolumn{2}{c}{$R<300$ kpc} \\
                 & pred. & obs.                   & pred. & obs. \\
  \hline
  $>2\times10^6$ & $11.6^{+5.7}_{-4.9}$ & $>$146 & $19.0^{+8.9}_{-7.5}$ & $>$193 \\
         $>10^7$ & $ 5.6^{+3.5}_{-2.9}$ & 16     & $ 8.5^{+5.0}_{-4.1}$ & 23     \\
         $>10^8$ & $ 1.1^{+1.5}_{-1.1}$ & 0      & $ 1.5^{+1.9}_{-1.4}$ & 0      \\
  \hline
\end{tabular}
\end{table}

The Fornax cluster has the best studied UCD population of any galaxy cluster \citep[e.g.][]{Mieske:2004, Gregg:2009} and therefore offers the best possibility to compare our predictions with observations. The observed number of UCDs in Fornax and their luminosities and for some cases dynamical masses are taken from a compilation of confirmed UCDs from work by \citet{Hilker:1999}, \citet{Drinkwater:2000}, \citet{Mieske:2002, Mieske:2004}, \citet{Bergond:2007}, \citet{Hilker:2007}, \citet{Firth:2007}, \citet{Mieske:2008}, \citet{Gregg:2009}, \citet{Schuberth:2010} and \citet{Chilingarian:2011}, where we assume UCDs are any objects with masses above $2\times10^6 \Msun$. The masses were calculated from the $V$-band luminosity and $(V-I)$ colour of the UCDs, assuming a $M/L_V$-$(V-I)$ relation according to \citet{Maraston:2005} SSP models (Kroupa IMF, blue horizontal branch) for ages above 11 Gyr \citep[see][]{MisgeldHilker:2011}.

Within a clustercentric radius of 0.9 degree, which corresponds to $\sim$300 kpc at the Fornax distance of 19 Mpc \citep{Ferrarese:2000}, the numbers of UCDs with masses above $10^7 \Msun$ are more than 95 per cent complete thanks to the all-targets approach by the 2dF Fornax surveys of \citet{Drinkwater:2000} and \citet{Gregg:2009}. For UCDs with masses above $2\times 10^6 \Msun$ the completeness is not easy to access. Most spectroscopic surveys concentrated on the inner 15 arcmin from the cluster center, i.e. a projected radius of $<$83 kpc at Fornax distance. Within 50 kpc the UCD number counts above $2\times 10^6 \Msun$ are nearly complete \citep[see discussion in][]{Mieske:2012}. Beyond this radius the spectroscopic coverage becomes patchy \citep[see][]{Schuberth:2010} and the completeness drops below 70 per cent within 100 kpc and probably is even lower beyond that. In terms of the total number the effect is not that dramatic, since the radial number density profile of UCDs decreases rapidly with clustercentric distance.

In Table \ref{tab:FornaxNumber} we show the predicted number of stripped nuclei for a simulated cluster with the virial mass of the Fornax cluster as well as the observed number of UCDs in the Fornax cluster. The predicted number of stripped nuclei are taken from equations \ref{eq:numFit1}-\ref{eq:numFit3}. We assume a virial mass for the Fornax cluster of $(7\pm2) \times 10^{13} \Msun$ \citep{Drinkwater:2001}. We predict about 1 stripped nucleus with a mass $M>10^8 \Msun$ in the Fornax cluster, which agrees well with the observed most massive `UCD3', which has a dynamical mass $M\approx 10^8 \Msun$ \citep{Hilker:2007}. The number of stripped nuclei with masses $M>10^7 \Msun$ we predict can account for 20 to 60 per cent of the UCDs in the Fornax cluster, with a mean of 40 per cent. For UCDs with masses $M>2\times 10^6 \Msun$ stripped nuclei can only account for 5 to 12 per cent of the number observed. This fraction is likely slightly lower due to the incompleteness of observed UCDs. If we assume all stripped nuclei are twice the effective radius of the original nucleus due to retaining stars from the host galaxy, the stripped nuclei will have a mass 1.5 times that of the nucleus \citep{Pfeffer:2013}. Taking this into account, we would predict $10.7^{+5.9}_{-4.9}$ stripped nuclei with masses $M>10^7 \Msun$. This would account for 25 to 70 per cent of the UCDs observed in Fornax with similar masses.

This means that most of the UCDs below $10^7 \Msun$ are most probably genuine globular clusters in accordance with the results of \citet{Mieske:2012}. For UCDs with masses above $10^7 \Msun$ a significant fraction are likely to be stripped nuclei.

\section{DISCUSSION} \label{sec:discussion}

\subsection{Are UCDs stripped nuclei?}

We find the contribution of stripped nuclei to UCD populations in galaxy clusters is only important for the most massive UCDs ($M>10^7 \Msun$). In the Fornax cluster stripped nuclei can only account for up to 26 per cent of UCDs more massive $2\times10^6 \Msun$. Therefore our finding suggests most UCDs are part of the bright tail of the GC population in agreement with \citet{Mieske:2012}.

Many UCDs appear to have elevated dynamical mass-to-light ratios, implying notable amounts of dark mass in them \citep{Mieske:2013}. \citeauthor{Mieske:2013} note that within their sample, two-thirds of high-mass UCDs ($M>10^7 \Msun$) and one-fifth of low-mass UCDs ($2\times10^6 < M/\Msun < 10^7$) require at the $1\sigma$ level some additional dark mass to account for their elevated dynamical mass-to-light ratio. They suggest central black holes as relict tracers of massive progenitors are a plausible explanation for the elevated mass-to-light ratios, which implies many of the more massive UCDs are stripped nuclei. Within the errors, these fractions are in good agreement with the number of UCDs we predict to form via tidal stripping in the Fornax cluster.

The difference between the observed number of UCDs and the predicted number of stripped nuclei suggests GCs dominate the combined GC plus stripped nucleus mass function, with the exception of the most massive objects. This agrees well with the luminosity function of UCDs in the Fornax cluster \citep[see fig. 4]{Gregg:2009} where the most luminous UCDs appear as a bright tail on top of the GC luminosity function. Interestingly, this is not seen in the GC mass function of \citet[see fig. 4]{Hilker:2009}. The origin of this difference is unclear. The difference in the slope of the power-law for GCs, $\alpha=-2$ for masses $M>3\times 10^5 \Msun$, and stripped nuclei, $\alpha=-1.4$ for masses $10^6 < M/\Msun < 10^8$, may provide a way to further test our tidal stripping model observationally. 

Other recent studies have investigated whether tidal stripping of dwarf galaxies in clusters can account for the UCDs observed in clusters. \citet{Mieske:2012} calculated the fraction of GCs that contribute to the UCD population based on the specific frequencies of GCs around galaxies. They found at most 50 per cent of UCDs can have been formed by tidal stripping. Around NGC 1399, they found if tidal stripping contributes 50 per cent of the observed UCDs than $\gtrsim 90$ per cent of primordial dwarf galaxies in the central $\sim 50$-70 kpc of the galaxy cluster must have been disrupted. However we find at most 12 per cent of UCDs are formed by tidal stripping. Repeating their calculation and taking this into account implies only $\sim 40$ per cent of primordial dwarf galaxies must have been disrupted. These calculations assume that any UCDs formed by tidal stripping must have formed at the centre of the galaxy cluster. In our model about 28 per cent of stripped nuclei associated with the central galaxy in a Fornax sized galaxy cluster have been stripped from satellite galaxies which have since merged in the cluster. This would then imply only $\sim 35$ per cent of primordial dwarf galaxies in the centre of galaxy clusters have been disrupted.

\citet{Thomas:2008} investigated a static model of tidal stripping in a Fornax-like galaxy cluster where dwarf galaxies are disrupted and form UCDs if they pass within a given radius. They found a static model predicts far too few UCDs at radii greater than about 30 kpc. Our model relieves some tension from the static models since not all UCDs must be formed in the cluster itself. Some UCDs are still associated with satellite galaxies within the clusters, while others have been stripped from satellite galaxies which have since merged in the cluster. We find approximately 36 per cent of stripped nuclei are associated with satellite galaxies in clusters. On average 22 per cent of stripped nuclei were formed around satellite galaxies that have since merged in the cluster. Stripped nuclei that are associated with satellites or formed around galaxies which later merged into the cluster will tend to be found at larger radii in clusters than those formed around the central galaxy and may therefore account for extended distribution of UCDs.

There are a number of further tests needed for our model. Comparing the absolute number of stripped nuclei predicted against the numbers observed is the most direct test. In particular comparisons of the number of stripped nuclei around central and satellite galaxies, as well as clusters of different masses, are particularly important. We find that stripped nuclei should scale with the halo virial mass for the galaxy. However, determining whether an object is a stripped nucleus or a normal GC is difficult without resolved stellar populations (such as in the Milky Way). Alternatively, enhanced mass-to-light ratios, above that expected from stellar populations, may indicate a tidal stripping origin rather than a GC origin \citep[e.g. if stripped nuclei host central black holes,][]{Mieske:2013}.  

We predict a minimum 20 per cent of stripped nuclei should be extended, with sizes more than twice that of the nucleus in the progenitor galaxy. It is likely the fraction is higher since interactions with satellite galaxies will have less affect on a dwarf galaxy than those with massive galaxy clusters. However, the observational completeness for such objects is relatively unknown and therefore more observations are required before a comparison can be made.

Since tidal tails of disrupting dwarf galaxies are expected to disperse and become unobservable on timescales of $\sim 1$ Gyr \citep{Pfeffer:2013} this gives constraints on the number of disrupting objects we can expect to observe in galaxy clusters. We find that an average of 0.3 objects (with stellar masses larger than $10^{7.5} \Msun$ before stripping) per $10^{13} \Msun$ have had mergers within the last 1 Gyr, or 0.25 per $10^{13} \Msun$ when only considering nucleated galaxies. Thus, given their halo virial masses, the Fornax and Virgo clusters are expected to have 2-3 and 11-14 disrupting galaxies which may be observable, and 2 and 9-12 disrupting nucleated galaxies, respectively.

\subsection{Caveats}

By far the largest source for error in our method is assuming that nuclei at high redshift adopt the same relations we observe at low redshift, in particular that the nucleus-to-galaxy mass ratio and the fraction of galaxies that host nuclei is constant at all times. If this is different in the early universe it may significantly affect our predictions for the number of stripped nuclei and their mass function. If nuclei are mainly formed after their progenitor galaxy this may imply a steeper mass function than we predict (Fig. \ref{plt:massFunction}) due to galaxies that merge early having less massive nuclei than similar galaxies which merge later. If at least some part of the nucleus is formed with the galaxy than the total number of stripped nuclei will be largely unaffected. If most nuclei are formed well after their host galaxy this could reduce the total number of stripped nuclei significantly. Some evidence for the latter exist in observations. \citet{Paudel:2011} find nuclei are typically much younger than their host galaxies (3.5 Gyr on average). However since the nuclei are modelled as simple stellar populations, any recent star formation may bias the nuclei to lower ages and does not rule out the formation of part of the nuclei at early times.

Other sources of error follow from the SAM. \citet{Weinmann:2011} find the dwarf-to-giant galaxy ratio in model clusters is too high by a factor $\sim$50 per cent. They suggest tidal disruption of low-mass galaxies is not efficient enough in the SAM. Since the ratio of disrupted to non-disrupted galaxies in the model clusters is about 60 per cent, this means we may be underestimating the number of stripped nuclei formed by 50 per cent. \citet{Guo:2011} note that the abundance of low-mass galaxies ($\sim 10^{10} \Msun$) is overproduced at early times ($z>0.6$), indicating that star formation in low-mass galaxies may be too efficient at early times in their model. This may not be a significant problem for our model because dwarf galaxies merge at much later times than when they form (see Fig. \ref{plt:mergerTimes}). However, this may affect the amount of time a nucleus has to form if it forms after the host galaxy and therefore relates to the first caveat. Low-mass galaxies in the SAM are also too strongly clustered on scales below 1 Mpc: too large a fraction of the model galaxies are satellites, although the overall abundance of galaxies matches well \citep{Guo:2011}. They suggest a lower value for the linear fluctuation amplitude $\sigma_8$ would reduce this clustering. This would affect our predictions little since the fraction of mass in a halo that was accreted in subhaloes of a given mass is relatively insensitive to the shape of the power spectrum \citep{Zentner:2003, Dooley:2014}. A lower value for $\sigma_8$ also results in subhaloes merging later which would allow more time for nuclei to form. There is only a slight improvement in these problems between MS-II and MSII-SW7 since the decrease in $\sigma_8$ between WMAP1 and WMAP7 is compensated by an increase in matter density $\Omega_m$ \citep{Guo:2013}.

In our model we don't take into account continuous tidal stripping of the stripped nuclei or that some objects may retain part of the host galaxy after stripping. Continuous tidal stripping will mainly affect the most massive and more extended objects near the centre of haloes (or objects with very radial orbits, but these will be fewer) and may need to be taken into account when comparing radial distributions with observations. This effect will mainly decrease high-mass stripped nuclei and therefore slightly increase the low-mass numbers (assuming low-mass stripped nuclei will tend to be more compact and therefore largely unaffected). During the tidal stripping process, objects that have few close passages in galaxy clusters may retain some part of their host galaxy \citep{Pfeffer:2013}. Objects that have many pericentre passages at distances larger than about 10 kpc may have an effective radius a factor of two larger than the isolated nucleus, or a factor of $1.5$ in mass (e.g. their simulations 4, 11 and 18), while objects that have many passages less than 10 kpc will appear about the same size as the isolated nucleus. Objects which have only 1 to 3 passages less than about 10 kpc, but many at much larger distances, may appear more than twice the size of the nucleus (i.e. their `box orbits'). In Section \ref{sec:orbits} we showed that the number of objects which have box orbits are likely to be few ($\sim 20$ per cent) and most objects will have many passages less than 10 kpc ($\sim 54$ per cent). Therefore for most objects retaining part of the host galaxy has little effect.

\section{SUMMARY} \label{sec:summary}

In this paper we present the first work to study GC and UCD formation within the framework of cosmology. We use cosmological simulations combined with a semi-analytic galaxy formation model to predict the properties of stripped nuclei that form via tidal stripping. Our main conclusions are summarized as follows:

\begin{itemize}
\item The number of stripped nuclei scales with cluster virial mass slightly less than linearly ($N \sim {M_\rmn{vir}}^{0.9}$): formation is slightly more efficient in low-mass clusters than high-mass ones. For individual galaxies it scales as $N \sim {M_\rmn{vir}}^{0.7}$. Stripped nuclei numbers only scale with galaxy stellar mass in the sense that high-mass galaxies typically reside in high-mass haloes, therefore there is large scatter. The average fraction of stripped nuclei associated with satellite galaxies in galaxy clusters is 35 per cent.

\item Between masses of $10^6$ and $10^8 \Msun$, the mass function of stripped nuclei is approximately a power-law with $N(M)\sim M^{-1.5}$. In order to compare the predicted mass function with observations we require observations that are complete to masses of a few times $10^6 \Msun$ for GCs and UCDs. We predict there should be a break in the combined GC and UCD mass function (although not at which mass) which changes from a GC dominated regime with $N(M)\sim M^{-2}$ to a stripped nucleus dominate regime, unless GCs completely dominate the mass function.

\item The progenitor galaxies of stripped nuclei are typically very old ($\sim 12$ Gyr), while the formation of stripped nuclei happens right up until 2 Gyr ago (our minimum time for formation). If nucleus formation happens, or continues to happen, some time between the formation and merging of the progenitor galaxy (within 1 Gyr for low-mass galaxies, 2 Gyr for high-mass galaxies with masses larger than $10^{10} \Msun$) the ages of stripped nuclei agree well with those of observed UCDs.

\item When distances are scaled by the cluster virial radius, the radial distributions of stripped nuclei in low-mass clusters are more concentrated than distributions in high-mass clusters. Detailed comparison with observed radial distributions is needed to further test our model, but is beyond the scope of this paper. Comparisons may be possible with programs such as the NGVS. 

\item During the formation of stripped nuclei, most objects have many close pericentre passages less than 10 kpc, while 20 per cent have only between one and three passages less than 10 kpc which is required for extended UCD formation (objects more than two time the size of the nucleus). Therefore tidal stripping will preferentially result in compact objects similar in size to the initial nucleus.

\item We predict that between 1 and 3 stripped nuclei more massive than $10^5\Msun$ and 1 to 2 stripped nuclei more massive than $10^6\Msun$ will form for systems with the virial mass of the Milky Way. For nuclei with masses larger than $10^6 \Msun$, this agrees well with the number of GCs in the Milky Way which have a spread in heavy element abundances and therefore were likely formed inside a dwarf galaxy. However for masses above $10^5 \Msun$ the number predicted is a factor of three lower than the number of GCs suggested to be stripped nuclei. The most massive nuclei predicted to form in the Milky Way and M31 agree well with the masses of $\omega$ Cen and G1, the most massive GCs from each galaxy, respectively.

\item In the Fornax cluster stripped nuclei can only account for up to 12 per cent of UCDs more massive than $2\times 10^6 \Msun$. For UCDs more massive than $10^7 \Msun$, between 20 and 60 per cent are likely to be stripped nuclei, or 25 to 70 per cent when taking into account stellar envelopes from tidal stripping. This agrees well with the result of \citet{Mieske:2012} that most UCDs are part of the bright tail of the GC population.
\end{itemize}

\section*{ACKNOWLEDGEMENTS}

We thank the referee, Thorsten Lisker, for helpful comments which improved the paper.

HB is supported by the Australian Research Council through Future Fellowship grant FT0991052 and 
Discovery Project grant DP110102608. 

The Millennium-II Simulation databases used in this paper and the web application providing online access to them were constructed as part of the activities of the German Astrophysical Virtual Observatory (GAVO).


\label{lastpage}

\end{document}